\documentclass{vgtc}                          

\ifpdf
  \pdfoutput=1\relax                   
  \usepackage{graphicx}                
  \DeclareGraphicsExtensions{.pdf,.png,.jpg,.jpeg} 
\else
  \usepackage{graphicx}                
  \DeclareGraphicsExtensions{.eps}     
\fi%

\graphicspath{{figures/}{pictures/}{images/}{./}} 

\usepackage{microtype}                 
\PassOptionsToPackage{warn}{textcomp}  
\usepackage{textcomp}                  
\usepackage{mathptmx}                  
\usepackage{times}                     
\usepackage{cite}                      
\usepackage{tabu}                      
\usepackage{booktabs}                  
\usepackage{paralist}

\usepackage{color}
\definecolor{RSTGuide}{RGB}{34,139,34}
\usepackage{xcolor}
\usepackage[normalem]{ulem} 
\newcommand\RSTGuide{\bgroup\markoverwith
  {\textcolor{RSTGuide}{\rule[-.5ex]{2pt}{2.5ex}}}\ULon}
\usepackage{scalerel}
\usepackage{setspace}

\onlineid{1198}

\vgtccategory{Research}

\vgtcinsertpkg

\title{Narrative Transitions in Data Videos}

\author{Junxiu Tang\\ %
    \parbox{1.4in}{\scriptsize \centering Zhejiang University\\ Hangzhou, China} 
\and Lingyun Yu\\ %
    \parbox{1.4in}{\scriptsize \centering Xi'an Jiaotong-Liverpool University\\ Suzhou, China} 
\and Tan Tang\\ %
    \parbox{1.4in}{\scriptsize \centering Zhejiang University\\ Hangzhou, China} 
\and Xinhuan Shu\\ %
    \parbox{1.4in}{\scriptsize \centering The Hong Kong University of Science and Technology\\ Hong Kong, China} 
\and Lu Ying\\ %
    \parbox{1.4in}{\scriptsize \centering Zhejiang University\\ Hangzhou, China} 
\and Yuhua Zhou\\ %
    \parbox{1.4in}{\scriptsize \centering Zhejiang Sci-Tech University\\ Hangzhou, China} 
\and Peiran Ren\\ %
    \parbox{1.4in}{\scriptsize \centering Alibaba Group \\ Hangzhou, China} 
\and Yingcai Wu\thanks{
        Yingcai Wu is the corresponding author.
        e-mail: ycwu@zju.edu.cn }\\ %
    \parbox{1.4in}{\scriptsize \centering Zhejiang University\\ Hangzhou, China}} 

\abstract{Transitions are widely used in data videos to seamlessly connect data-driven charts or connect visualizations and non-data-driven motion graphics.
To inform the transition designs in data videos, we conduct a content analysis based on more than 3500 clips extracted from 284 data videos.
We annotate visualization types and transition designs on these segments, and examine how these transitions help make connections between contexts.
We propose a taxonomy of transitions in data videos, where two transition categories are defined in building fluent narratives by using visual variables.

} 

\CCScatlist{
  \CCScatTwelve{Human-centered computing}{Visu\-al\-iza\-tion}{Visu\-al\-iza\-tion theory, concepts and paradigms}{}
}

\begin{document}


\begin{spacing}{0.965}
\firstsection{Introduction}

\maketitle
Recently, videos incorporated with data-driven facts and motion graphics, also known as \textit{data videos}, have become a popular medium for telling data stories~\cite{Amini2015,Amini2017,Tang2020}.
With data-driven visual expressions, rich information can be packed in a data video and organized in a sequence of story pieces.
Transition, which refers to the changes among different ``informationally-distinct'' story pieces, plays an important role in creating an engaging experience for audiences\cite{Hullman2013}.

\begin{figure}[htb]
  \setlength{\abovecaptionskip}{5pt}
  \setlength{\belowcaptionskip}{-12pt}
  \centering
  \includegraphics[width=0.9\columnwidth]{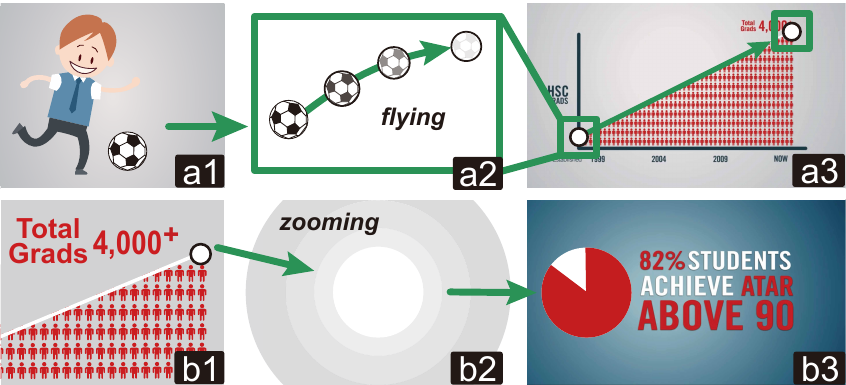}
  \caption{
      Screenshots of a data video\cite{Matrix2012} that shows two types of transitions, namely, between non-visualization and visualization content (a1-a3), and between two visualizations (b1-b3).
  }
  \label{fig:intro-example}
  \end{figure}

Transition is an effective way to cement story pieces together.
First, transitions are identified as a major type of attention cues in data videos, which helps engage and orient viewers \cite{Amini2015}.
In addition, previous research has shown that transitions can help viewers perceive the relationship between data for comparison~\cite{Ondov2018} or aggregation~\cite{Kim2019}.
Moreover, transitions, by linking different contents in a linear sequence, allow audiences to comprehend the narrative connections better \cite{Xia2018}.
For example, as shown in \autoref{fig:intro-example}, in the advertisement for an education association \cite{Matrix2012}, the designer uses a line chart to illustrate the increasing number of their customers, and a pie chart to show the contribution of the professional services.
The video segment has two types of transitions.
First, in \autoref{fig:intro-example}(a1), a little boy kicks away a football, which is then transformed into the start point of the line (\autoref{fig:intro-example}(a2)) and grows along with the line chart (\autoref{fig:intro-example}(a3)).
The animation on the football establishes a transition from the non-visualization content (\autoref{fig:intro-example}(a1)) to the visualization content (\autoref{fig:intro-example}(a3)). 
Second, in \autoref{fig:intro-example}(b1), with the camera zooming into the circle at the end of the line (\autoref{fig:intro-example}(b2)), the circle is transformed into a pie chart that shows the percentage of high-score students (\autoref{fig:intro-example}(b3)). 
The transformation of the circle establishes a transition which connects the line chart and the pie chart, thereby indicating that the latter pie chart is showing the distribution of the end year on the line chart.

The aforementioned case provides a vivid example that transitions can help articulate contents in data videos.
A rich palette of transition designs has emerged in practice.
On the one hand, designing the transitions between visualizations has been an ongoing effort of the visualization community.
For example, Heer et al.~\cite{Heer2007} proposed staged animation to transform between statistical charts.
Hullman et al.~\cite{Hullman2013} examined a set of transition types and gained a deeper understanding of narrative sequences.
Amini et al.~\cite{Amini2017} summarized staged animations~\cite{Heer2007} in data videos as one type of transition clips in the taxonomy of data clip types.
However, they did not provide a comprehensive analysis on how transitions are designed.
Moreover, their designs for transitions are relatively unitary.  
On the other hand, video designers attach great importance to the detailed transition designs \cite{Richardson2020,Xia2018}.
Nevertheless, these designs mainly focus on motion graphics and are not much related to visualizations.
The investigation of transition, such as visual designs and effects on expressing narrative information, can help designers and visualization researchers create attractive data stories.
However, a taxonomy of animated transitions in data videos not only between statistical charts but also between visualizations and other narrative animated elements (e.g., icons and texts) is still missing.

To address this issue, we investigate the transition design of visualizations in data videos.
We study the transition designs in 284 data videos created by professional designers from a narrative perspective and conduct a content analysis to explore the design patterns.
We focus on the transitions not only between two visualizations but also between visualizations and other video contents.
Elaborate transition designs in these videos provide abundant and valuable cases on how transitions help articulate the narrative. 
Informed by the survey and related literature\cite{Heer2007,Segel2010,Hullman2013,Amini2015,Amini2017}, we propose a taxonomy about transition designs in data videos.
We identify five types of typical narrative transitions, two of which can preserve the context during the transition period, namely,~\textit{Preserving Guide} and \textit{Narrative Agent}. 


\section{Related Work}
In this section, we introduce related work with regard to animated transitions and data videos.
Animated transitions are commonly used in visualization presentations.
Heer and Robertson\cite{Heer2007} examined the effect of animated transitions between statistical charts and proposed detailed designs concerning the congruence of contexts.
On the basis of an evaluation of those designs, they indicated that some animated transitions could keep viewers oriented.
This notion means that well-designed transitions between visualizations can improve the viewer's perception and cognition.
Researchers have recently conducted more specific work on visualization transition designs for different presentation tasks, such as the analogy between visualizations~\cite{Ruchikachorn2015}, aggregation operations\cite{Kim2019}, and visual grouping\cite{Chalbi2020}.
As for design details, Thompson et al.\cite{Thompson2020} proposed several dimensions of animated data graphics and classified transitions based on these dimensions.

Researchers also used animated transitions for data-driven storytelling.
In Segel and Heer's design space, as a type of visual narrative tactic, \textit{transition guidance} promotes the use of conventional methods to achieve the continuity among different shots to keep viewers oriented when old scenes destruct and new scenes come\cite{Segel2010}.
Animated transitions can be a type of transition guidance when telling stories.
Hullman et al.~\cite{Hullman2013} investigated the arrangements of narrative sequences based on transitions.
Furthermore, Kim et al.~\cite{Kim2017} developed GraphScape, a synthetic model to evaluate the transition cost of visualization sequences.
Researchers also use transitions to make stories cohesive in timelines~\cite{Brehmer2016} and slideshows~\cite{Wang2018}.
Amini et al.~\cite{Amini2017} include \textit{Transition} in the taxonomy of data clip types.
They expanded a few transition designs in their proposed authoring tool, DataClips, from and to pictographs based on Heer's work~\cite{Heer2007}.

However, the discussed transition designs in the previous work have not well covered all transitions that connect contexts in the narrative of data videos.
Data videos have a series of different forms, for example, animated graphics that consist of icons and characters, standard charts, pictographs.
Given that not only visualizations exist in the data videos, but also iconic motion graphics, transition designs between scenes have a large space.
In this paper, we expand the taxonomy to include specific designs and usage scenarios of transition clips.
Additionally, we pay attention to how they cement the context in the linear narrative in data videos.
\section{Content Analysis}
We first state the definition of a transition in data videos in our research scope, followed by the adopted methodology.


\subsection{Definition}
Prior work has defined transitions in narrative visualizations\cite{Hullman2013}.
Hullman et al.~\cite{Hullman2013} consider the change between two independent visual expressions as a transition. 
In the previous investigation of data videos, Amini et al. \cite{Amini2017} find that ``most transitions in these videos are a combination of destroy/create clips rather than staged animations described in~\cite{Heer2007}''.
However, after collecting and analyzing 284 data videos, we found many more different transition designs in data videos compared with the previous work.
For example, transitions in \autoref{fig:intro-example} connect the contents by using the shared visual elements of the successive scenes.
The design of using shared visual elements is not strictly the staged animation~\cite{Heer2007}, but they have similarities in reusing visual elements and keeping audience oriented in syntax and in semantic.
In our work, we follow the prior definition of transitions ~\cite{Heer2007,Hullman2013,Amini2017} and expand it in a wider scenario.
First, we identify \textit{clips} in data videos, where a clip is an elemental unit of the data video sequence~\cite{Amini2017}. 
Each clip is considered containing two narrative states and one transition between these two states.
A \textit{narrative state} in a clip is defined as an informationally-distinct scene for presenting data facts or other video narrative, following the definition of \textit{narrative visualization state} ~\cite{Hullman2013}.
Therefore, we define the \textit{transition} (also called \textit{narrative transition}) in our paper as the change of two narrative states in a clip, where the narrative state can be 
\textit{1)} an animated content without visualizations or data-driven arguments; 
or \textit{2)} a visualization content that includes standard charts, pictographs, and other data-driven arguments.

\subsection{Methodology}
We conducted a content analysis~\cite{Krippendorff2018}, which was also used in previous work ~\cite{Segel2010,Hullman2011,Byrne2015} to gain a comprehensive understanding of the transition design in data videos.

First, we collected 284 data videos from reputable sources, such as the leading media outlets, design associations’ portfolios, and video sites.
We gathered these videos following the same criteria in Amini’s work \cite{Amini2017}.
The videos should convey data-driven arguments and contain related visualizations.
The wide range of topics in the videos includes science, finance, politics, sports, and history.
The video dataset includes animated motion graphics, photography videos, and combinations of them.
We took apart these videos into single clips.
We only focused on the clips that have at least one scene with visualizations or data-driven arguments. 
Finally, our dataset contains 3909 clips: 1644 vis-to-vis clips (42.1\%), 1104 others-to-vis clips (28.2\%) and 1161 vis-to-others clips (29.7\%). 

To examine the detailed transition designs in clips, we first reviewed selected samples and proposed an initial taxonomy considering the changing visual variables (e.g., position, color, and shape) ~\cite{Bertin1983} during the transition. 
We iteratively improved the taxonomy through multiple rounds of discussion among authors and attempts to label sampled transitions. 
Ultimately, we achieved an agreement on the final taxonomy (Sec. \ref{section:taxonomy}).
We mainly considered the following questions when analyzing the transition design:
\begin{compactitem}[$\bullet$]
    \item What are the narrative states (e.g., visualizations, animated icons) of the transition?
    \item What visual variables have changed, and what unchanged during the transition? 
    \item How are the transitions visually presented?
    \item What is the narrative relationship between two states?
\end{compactitem}
Based on the taxonomy, two authors independently coded all the clips and reached an initial consensus on 87.7\% (3430) transitions. 
The conflicts (12.3\%, 479) were further resolved through discussion.
Based on the agreement, we conducted a quantitative analysis on the visualization types and transition types in each clip.
The complete results are attached to the supplemental material.
  


\section{Taxonomy of Transitions} \label{section:taxonomy}
This section first presents an overview of our proposed taxonomy for narrative transitions in data videos and then introduces data-driven transitions in detail.

\subsection{Overview}
\begin{figure*}[htb]
\setlength{\abovecaptionskip}{3pt}
\setlength{\belowcaptionskip}{-12pt}
\centering 
\includegraphics[width=0.85\textwidth]{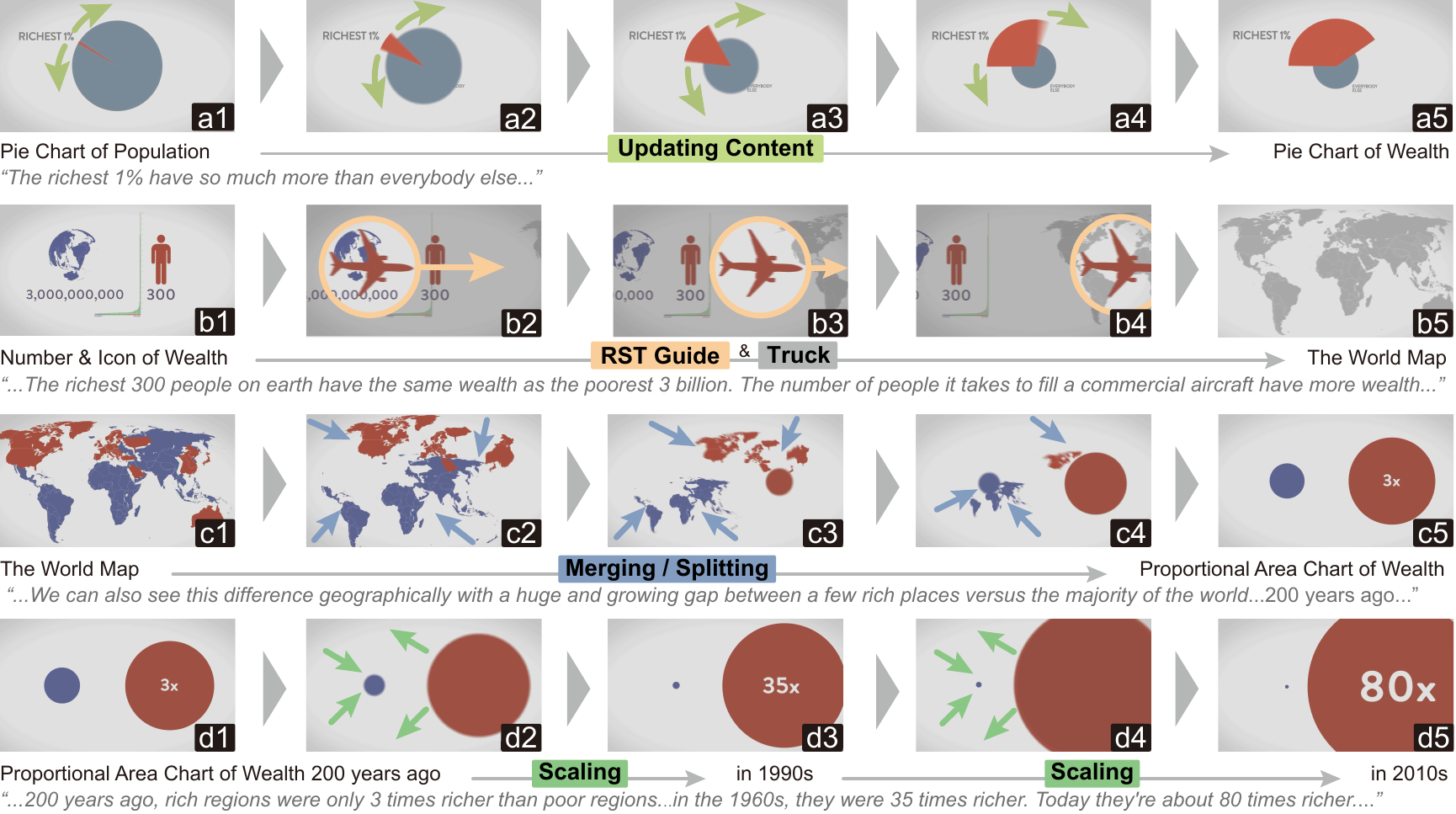}
\caption{
    Screenshots of the video \textit{Global Wealth Inequality}\cite{TheRulesOrg2013}.
    We added green marks on the screenshots to show the animation between two consecutive scenes.
    We presented the transition types, visual content and the corresponding transcripts (gray text) under each clip sequence. 
}
\label{fig:case-1}
\end{figure*}

According to the definition of narrative transitions (Sec. 3.1), we examine the designs from two aspects, namely,  
data-driven animated transitions (e.g., staged animations~\cite{Heer2007}, \autoref{fig:intro-example}(b), and \autoref{fig:case-1}(a)(c)(d)) and non-data-driven ones (e.g., fade-in/off, \autoref{fig:intro-example}(a), and \autoref{fig:case-1}(b)).
We list five identified transition types in an order of the relevance to data, where the first three pertain to non-data-driven transitions and the last two belong to data-driven ones.
The frequency of each transition type in the dataset (3909 clips) is also reported.
Note that multiple types can be combined together to establish a transition in a clip. 
For example, \autoref{fig:case-1}(b) uses a camera motion and a preserving guide in the case. 
We provide animated illustrations of these narrative transitions on the website: \url{https://narrativetransitions.github.io/home/}.

\begin{figure}[htb]
\setlength{\abovecaptionskip}{5pt}
\setlength{\belowcaptionskip}{-10pt}
\centering
\includegraphics[width=0.9\columnwidth]{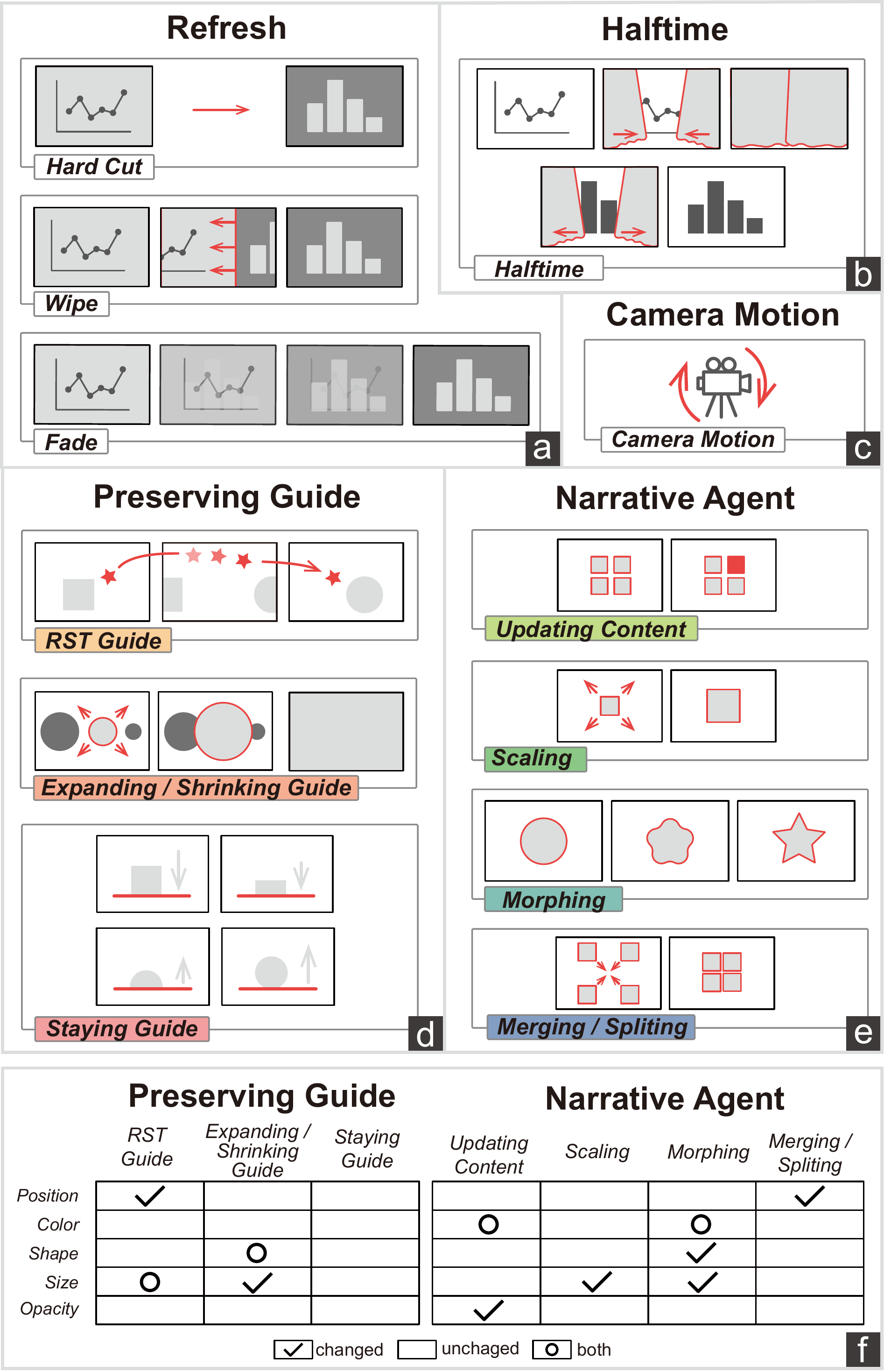}
\caption{
    Taxonomy of transition designs in data videos (a-e).
    We also present changed and unchanged visual variables of \textit{Preserving Guide} and \textit{Narrative Agent} transitions (f).
    }
\label{fig:taxonomy}
\end{figure}

\textbf{Refresh} (53.0\%, 2070) means a complete update of the previous scene (\autoref{fig:taxonomy}(a)).
In this type of transitions, no connection exists between the last frame of the previous scene and the first frame of the coming scene.
Usually, this mechanism can be used to present a new topic or an abrupt turn.
We place the combinations of the destruction of previous scenes and creation of coming scenes in this category, for example,~\textit{Hard Cut},~\textit{Fade}, and \textit{Wipe}.

\textbf{Halftime} (2.5\%, 98) adds a new scene between video context (\autoref{fig:taxonomy}(b)).
Such a scene is comparable to a quick half time or the stage curtain between the previous scene and the coming scene.

\textbf{Camera Motion} (14.9\%, 583) updates the scene due to the changing viewpoints or screen focus.
The seven subtypes of camera motion are\cite{Rea2015,Storyblocks2019}: \textit{Pedestal}, \textit{Truck}, \textit{Tilt}, \textit{Pan}, \textit{Dolly}, \textit{Zoom}, and \textit{Rack Focus}.
\textit{Pedestal} means moving the camera vertically.
By contrast, \textit{Truck} means moving the camera horizontally.
\textit{Tilt} and \textit{Pan} also mean moving the camera in the vertical and horizontal direction, respectively.
However, they both require the camera to keep a stable focus during the movement.
\textit{Dolly} means moving the camera forward or backward.
\textit{Zoom} means changing the focal distance.
\textit{Rack Focus} means changing the focus in the scene, for example,  bokeh effects.
Camera movements can create a sense of space over scenes, and usually, they are used to present spatial visualizations, such as maps.
The changes in the focus of scenes can highlight the key points of narrations according to the video designer's intention.

\textbf{Preserving Guide} (22.9\%, 894) reuses elements in the previous scene as a visual guide that directs to the next scene (\autoref{fig:taxonomy}(d)).
For example, a preserving guide could be a flying icon (\autoref{fig:case-1}(b)), a colored area, or a stable line, to lead the viewers' attention between two consecutive scenes.
This guide can be used with camera motions to construct fluid transformations (\autoref{fig:case-1}(b)).

\textbf{Narrative Agent} (19.9\%, 777) means substitutes for data during data-driven storytelling (\autoref{fig:taxonomy}(e)).
The transitioned elements could be regarded as agents of data attributes or data values.
This transition illustrates the change of data such as \textit{Scaling} and \textit{Merging}.



Different transition designs are useful to connect diverse narrative states.
For example, in bar charts, data-encoded bars can be used as \textit{Narrative Agents} to present the change of data.
Another example is that \textit{Camera Motion} can be used to transform among different places on maps because this transition can generate a sense of space.

\vspace{-5pt}

\subsection{Data-driven Narrative Transition}
Transitions, such as \textbf{Refresh}, \textbf{Halftime} and \textbf{Camera Motion}, can be used not only in data videos but also in other video narratives.
\textbf{Refresh} is the most popular transitions in the dataset.
This kind of transition has been integrated in existing video authoring tools~\cite{Premiere2020,AfterEffects2020} and can be easily employed in videos.
\textbf{Preserving Guide} and \textbf{Narrative Agent} need craft animations to conduct fluent narrative.
These two transitions are usually used in data videos to preserve data attributes or encoding data-driven insights, which present narrative information of the visualization content of data videos.

We paid special attention to \textbf{Preserving Guide} and \textbf{Narrative Agent} transitions to understand the data-driven design patterns of transitions in data videos.
They make up a large proportion of the studied clips, 
especially in the 1644 vis-to-vis transitions, among which,
477 are \textbf{Preserving Guide} (29.0\%) and 617 are \textbf{Narrative Agent} (37.5\%).
Within these transitions, changed and unchanged visual elements both exist.
We focus on the visual contents within clips in terms of Bertin's variables of the image\cite{Bertin1983}.
Those changed variables consist of the animation of transitions, while unchanged ones preserve the narrative information to maintain the congruence of the previous scenes and coming scenes.
The narrative information could be data attributes and values preserved in special visual variables.
The constant variables, which are similar to data agents, help viewers understand what narrative is going on.
We list them in \autoref{fig:taxonomy}(f).
Note that these agent elements are not isolated from each other in practice.
In the following sections, we introduce the narrative transitions of data videos in detail.
Illustrations about each type of transition are presented in \autoref{fig:taxonomy}(d) and (e).

\subsubsection{Preserving Guide}
\includegraphics[width=16em, trim=0 0.45em 0 0]{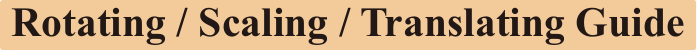}
A single or a group of visual items in the previous scene are kept and reused in the coming scenes.
However, the layout of these visual items, for instance, orientation, size, and position, may change.
These visual elements maintain the narrative information in previous scenes by preserving the same appearance, and at the same time, guide viewers' attention to a new scene by changing the layout.
This transition is particularly useful in setting up a supplement, correlation, or a comparison.
We abbreviate this transition as \textbf{RST Guide}.

\includegraphics[width=12.5em, trim=0 0.45em 0 0]{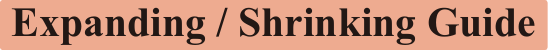}
This type means two related situations.
The first situation is to expand the color of a specific item in the previous scene to the background color of the coming scene.
Often, the coming scene introduces details about the item.
The other situation is to shrink the background color of the previous scene to a specific item of the coming scene.
This type of transition is usually used when presenting the relation between the whole and the parts.
Color is an essential visual variable in this type of transition: the same color of adjacent scenes illustrates the same subject.

\includegraphics[width=6.5em, trim=0 0.45em 0 0]{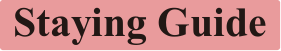}
This type of transition means that the layout of the visual items in the previous scene remains the same, and additional items are interacting with the existed items.
The stable visual items are unchanged cue in the scene, and they can be regarded as the basis for the additional items entering, growing, and leaving.

\subsubsection{Narrative Agent}
\includegraphics[width=8em, trim=0 0.45em 0 0]{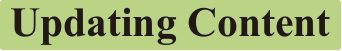}
This type of transition means updating the visual content without changing the shapes, positions, and colors of the contents in the scenes.
In this transition, the shape, position, and color maintain the narrative information; however, the number of the contents changes to show the differences.
Such a mechanism is used in data videos for highlighting or presenting data sequentially.

\includegraphics[width=3.7em, trim=0 0.45em 0 0]{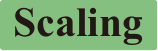}
These transitions change sizes to show the value change.
Although the size changes, the other visual variables remain the same, thereby making the contents coherent in different scenes.

\includegraphics[width=4.5em, trim=0 0.45em 0 0]{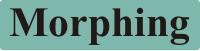}
This type of transition mainly morphs icons or shapes from old ones to new ones.
This transition conveys insights into the transformation between data.

\includegraphics[width=8em, trim=0 0.45em 0 0]{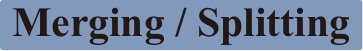}
Merging means combining separate contents into a group, while splitting means separating a group of contents.
This type of transition describes insights into the gathering or scattering. 
The whole-part relationship of data can be clearly illustrated by using this type of transition.
\section{Case Study of Narrative Transitions}

In this section, we take clips from one data video as a case to illustrate the effect of transitions in connecting narratives and visualizations.
We highlight characteristic clips of these videos and mark transitions in \textbf{bold} texts.
This video is about global wealth inequality\cite{TheRulesOrg2013}.
It presents a series of data facts about the wealth comparison between the poor and the rich.
Most facts are linked through crafted transitions rather than simply fading in/off.
We introduce five clips (\autoref{fig:case-1}) in detail.

The story starts with a pie chart, which shows the population distribution (\autoref{fig:case-1}(a1)).
The red color represents the richest people, which is only 1\%, 
while the pewter represents the other 99\% of the people.
Then, the pie chart \includegraphics[width=8.5em, trim=0 0.45em 0 0]{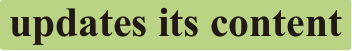} to present another data fact (\autoref{fig:case-1}(a2-a5)), where the thin red sector expands, while the pewter part reduces.
By establishing contrasts of wealth and popularity between two groups of people, this transition reveals that the 1\% richest people own much more than the rest.
Within the transition, the color encodes the data category (the richest people and the rest).
The color encoding also preserves the narrative information during the transitions of two pie charts to keep viewers oriented.

The following footage consists of two icons and texts  (\autoref{fig:case-1}(b1)).
The scene shows the fact that 3 billion poorest people and 300 richest people own the same wealth.
Afterward, the narrator mentions that the number of people it takes to fill a mid-size commercial aircraft has more wealth than the combined populations of
four countries.
In the clip, an aircraft icon flies through the scene, and the previous pictograph quits the scene under the \textbf{Truck} transition (\autoref{fig:case-1}(b2-b4)).
The icon moves from left to right and guides the viewers' attention to the following world map.
After this contextual data fact, the topic turns from inequality over the population to that over regions (\autoref{fig:case-1}(b5)).
The aircraft in this clip is a \includegraphics[width=5em, trim=0 0.5em 0 0]{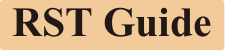} that attracts viewers' attention and shifts the presented topic insensibly.

Then, the world map splits into two parts by two colors (\autoref{fig:case-1}(c1)).
Developed regions and less developed regions are encoded with red and blue, respectively.
These regions first \includegraphics[width=2.4em, trim=0 0.5em 0 0]{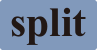} from the whole world map and then \includegraphics[width=3em, trim=0 0.4em 0 0]{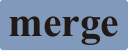} into two different circular areas (\autoref{fig:case-1}(c2-c4)).
Each circular area represents the total wealth of the merged regions.
The areas consist of a proportional area chart of global wealth measured by poor and rich regions.
The chart then \includegraphics[width=8.5em, trim=0 0.4em 0 0]{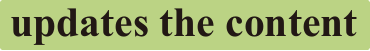} by \includegraphics[width=3.5em, trim=0 0.45em 0 0]{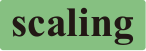} to show the comparison of the rich regions and the poor regions over two hundred years (\autoref{fig:case-1}(d1-d5)).
During this process, the layout and the color of the two circles keep constant, whereas the sizes of the two circular areas change according to the data.
The change in sizes creates a prominent contrast on the wealth of the rich and poor regions.

\section{Discussion and Conclusion} \label{section:discussion-conclusion}
In this study, we investigate the taxonomy of transition designs in data videos.
First, we collect the dataset of 284 professional data videos with 3909 clips.
These videos cover various visual styles of visualization contents and diverse transition designs.
Based on a content analysis on these clips, we propose a taxonomy of narrative transitions in data videos.
With regard to the change of visual variables, we conduct a more in-depth analysis of data-driven narrative transitions that preserve narrative meanings of contents, namely, \textit{Preserving Guide} and \textit{Narrative Agent}.

The proposed taxonomy for narrative transitions takes a step in this direction and hopes to encourage future research.
First, following the taxonomy, 
evaluations on specific transition designs are needed to assess the effectiveness of engagement and memorability of data videos.
Designers can take transition designs into consideration when building the attention cues of data videos, for example, highlighting data facts or guiding viewer's attention.
Second, our taxonomy provides a new way of inspecting the relationship between narrative transitions and story sequences in data videos.
Narrative transitions can not only highlight visual changes in presentations but also enrich narrations.
We plan to propose a comprehensive model that considers both narrative transitions and story sequences.
Third, our taxonomy can inspire the transition design for other genres of narrative visualizations, for example, animated long-form web articles or data-GIFs\cite{Shu2021}.
Finally, except for narrative, future work can consider other messages that transition can present, for example, visualization rhetoric\cite{Hullman2011} and acquire codes\cite{Byrne2015}.
\end{spacing}

\acknowledgments{
The authors wish to thank reviewers for their valuable comments and suggestions. 
The work was supported by NSFC (61761136020), National Key R\&D Program of China (2018YFB1004300 ), NSFC-Zhejiang Joint Fund for the Integration of Industrialization and Informatization (U1609217), Zhejiang Provincial Natural Science Foundation (LR18F020001) and the 100 Talents Program of Zhejiang University. This project was also partially funded by Microsoft Research Asia. L. Yu is supported by XJTLU Research Development Funding RDF-19-02-11.
}

\bibliographystyle{abbrv-doi}

\newpage
\bibliography{reference}
\end{document}